\documentstyle[preprint,aps]{revtex}
\input psfig

\def\xvp{{\bf x}^{\prime}}

\def\xv{{\bf x}}
\def\kv{{\bf k}}
\def\qv{{\bf q}}
\def\Ha{{\cal H}}
\def\Rv{{\bf R}}
\def\ut{{\tilde u}}
\begin{document}
\draft
\title{Kosterlitz-Thouless Transitions on Fluctuating Surfaces}
\author{Jeong-Man Park and T.C. Lubensky}
\address{Department of Physics, University of Pennsylvania, Philadelphia
PA 19104}
\maketitle
\begin{abstract}
We investigate the Kosterlitz-Thouless transition for hexatic order on a
free fluctuating membrane and derive both a Coulomb gas and a sine-Gordon
Hamiltonian to describe it.  In the former, both disclinations
and Gaussian curvature contribute to the charge density.  In the latter,
there is a linear coupling between a scalar field and the Gaussian
curvature.  We derive renormalization group recursion relations that
predict a transition with decreasing bending rigidity $\kappa$.
Using the sine-Gordon model, we show that there is no KT transition on a
deformable sphere.
\end{abstract}
\par
A fluid membrane can be modeled as an ideal fluctuating surface whose
energy depends only on its geometry.
If the membrane is free from external tension, 
the dominant term in this energy is the Helfrich-Canham
bending energy\cite{Helfrich73-1,Jerusalem89}
with bending rigidity $\kappa$. Such a
membrane with a molecular length $a$ is crumpled at length scales beyond the 
persistance length, $\xi_p = a e^{4 \pi \kappa /3T}$,
at which the renormalized bending rigidity
passes through zero\cite{Helfrich85-1}.
Membranes with hexatic tangent plane orientational order are stiffer than
fluid membranes\cite{Nelson87-1}.  A flat membrane has quasi-long-range
(QLR) hexatic order at low temperature and undergoes a Kosterlitz-Thouless
(KT) disclination unbinding transition to a disordered
state\cite{Nelson79-1}.  A free hexatic membrane at low temperature
exists in a ``crinkled" state with QLR hexatic order and 
a nonzero long-wavelength bending rigidity\cite{David87-1,GuitKardar90-1}.
\par
Free hexatic membranes are predicted\cite{GuitKardar90-1} to undergo a
KT transition from the crinkled phase to the crumpled fluid phase.  To
our knowledge, however, there has been no systematic treatment of this
transition with complete  renormalization group recursion relations
for the hexatic stiffness $K$ and disclination fugacity $y$ associated
with the KT transition in addition to 
the bending rigidity $\kappa$ controlling 
membrane height fluctuations.  
In this paper, we present an analysis of the KT transition on a fluctuating
membrane using both the Coulomb gas model obtained directly from the
hexatic Hamiltonian and the sine-Gordon model dual to the Coulomb gas
model. A more detailed presentation will appear in a separate
publication.
\par
\begin{figure}[h]
\centerline{\psfig{figure=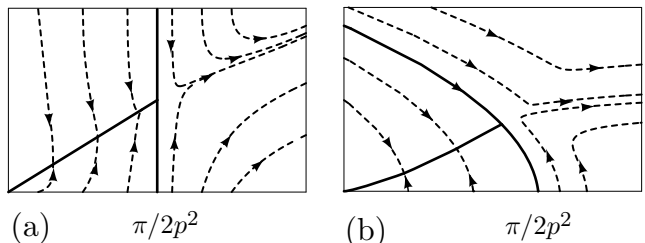}}
\caption{Renormalization flows obtained from Eqs.\ (\protect\ref{nrecur1})
and (\protect\ref{nrecur3}) in (a) the $(x,z)$ or 
$((\beta {\overline K})^{-1}, (\beta \kappa )^{-1})$-plane 
and (b) in the $((\beta K)^{-1}, (\beta \kappa )^{-1} )$-plane.  (b) shows
that increasing either $(\beta K)^{-1}$ or $(\beta \kappa )^{-1}$ leads to
melting of the crinkled phase.}
\label{fig1}
\end{figure}

Both local Gaussian curvature and disclinations contribute to the charge
density of the coulomb gas.
The bare hexatic rigidity $K$ plays the role of 
the square of one of the charges in the Coulomb gas.  
The other charge is determined by $K$ and the Liouville
action\cite{David87-1}. As in electrodynamics, neither charge 
undergoes a renormalization in agreement with Ref. \onlinecite{David87-1},
which found that $K$ did not renormalize.
We find, however, that fluctuations in Gaussian
curvature reduce the dielectric constant even in the absence of
dislcinations and lead to a decrease with decreasing $\kappa$
of the effective hexatic rigidity that controls correlations of the hexatic
order parameter. 
As a result, we find, in contrast to previous
treatments\cite{David87-1,GuitKardar90-1} that 
decreasing $\kappa$ leads to a KT transition from
the crinkled to the crumpled state as depicted in Fig.\ \ref{fig1} .  We obtain
recursion relations for $K$, $y$, and $\kappa$  
that yield a KT transition and that allow us
to calculate quantities such as the persistence length 
in the crumpled phase.    
We find that  
$\xi_p \sim \xi_{\rm KT} e^{4 \pi \kappa /3 T}$
where $\xi_{\rm KT} \sim a \exp(\pi/b\sqrt{T-T_c})$ is the KT correlation
length when $\kappa$ is large and the persistance length is larger than the
$\xi_{KT}$.  
\par
The sine-Gordon model we derive applies to an open membrane and to closed
membranes (vesicles) of arbitrary genus $h$.  As on flat surfaces, 
this model has a scalar field $\phi$ with a gradient and a $\cos \phi$
energy.  The principal coupling to geometry occurs via an interaction,
similar to the dilaton coupling of string theory\cite{dilaton},
proportional to $\phi R$ with an {\it imaginary} coefficient, where $R$ is
the scalar or intrinsic curvature.  
Using standard renormalization group techniques
for the sine-Gordon model\cite{AmitGoldGrin80}, we obtain the same
recursion relations as we obtain using the Coulomb gas.  We further show
that shape fluctuations on a sphere, like gauge fluctuations in an infinite
two-dimensional superconductor, generate a ``mass" for $\phi$ and
depress the KT transition temperature to
absolute zero.
\par
Membrane coordinates in three space are specified by a vector ${\bf R} ( {\tilde u} )$
as a function of a two-dimensional parameter ${\tilde u} = ( u^1 , u^2 )$.  The
metric tensor $g_{ab} = \partial_a {\bf R} ( {\tilde u} ) \cdot \partial_b \Rv ( \ut
)$, its inverse $g^{ab}$, and the curvature tensor 
$K_{ab} = {\bf N} \cdot \partial_a \partial_b
{\bf R}$, where $\partial_a = \partial/\partial u^a$ and $\bf N$ is the unit
normal to the surface, can be constructed from ${\bf R}$.  
The energy of a free membrane is described by the Helfrich-Canham
Hamiltonian\cite{Helfrich73-1}
\begin{equation}
{\cal H}_{\kappa} = \case{1}{2} \kappa \int d^2 u \sqrt{g} H^2 ,
\end{equation}
where $g = \det g_{ab}$ and $H = g^{ab}K_{ab}$ is 
twice the mean or extrinsic curvature.
A membrane with hexatic order has an order parameter $\psi = |\psi|e^{6 i
\theta}$.  
A membrane with generalized 
$p$-atic order has an order parameter $\psi = | \psi | e^{i
p \theta}$ invariant under $\theta \to \theta + (2 \pi /p )$.  
The elastic energy associated with spatial variations of $\theta$ 
for system with $p$-atic order is\cite{Nelson87-1}
\begin{equation}
{\cal H}_{\theta}  =  \case{1}{2} K \int d^2 u \sqrt{g} g^{ab} (\partial_a
\theta - A_a ) ( \partial_b \theta - A_b ) ,
\end{equation}
where $A_a$ is the spin connection satisfying
$\gamma^{ab} \partial_a A_b = S$, where $\gamma^{ab}$ is
the anitsymmetric tensor with magnitude $g^{-1/2}$, 
and $S = \det K_a^b = R/2$ is the Gaussian curvature.
The total long wavelength Hamiltonian for a $p$-atic
membrane is then ${\cal H}= \Ha_{\kappa} + \Ha_{\theta}$. 
In the presence of disclinations, which have strength $q = k/p$ where $k$
is an integer, $\partial_a \theta$ becomes singular: 
$\partial_a \theta =
\partial_a {\tilde \theta} + v_a$. 
The singular part satisfies $\gamma^{ab}
\partial_a v_b = n$, where
$
n = 2 \pi g^{-1/2} \sum_{\alpha} q_{\alpha} \delta ( {\tilde u} -
{\tilde u}_{\alpha} ) 
$
is the disclination density arising from disclinations of strength
$q_\alpha$ at positions ${\tilde u}_{\alpha}$.  
Thus $- \Delta_g ( v_a - A_a ) =
\gamma_a{}^b  \partial_b ( n - S )$, where $\Delta_g = D_a D^a$
is the covariant Laplacian operator on a surface with metric tensor
$g_{ab}$.  Using this relation, ${\cal H}_{\theta}$ can be
expressed as the sum of a regular part depending only on $\tilde \theta$
and a curved space generalization of the energy of a $2D$ Coulomb gas: 
${\cal H}_{\theta} = \Ha_{\rm reg} + \Ha_c$ with
\begin{eqnarray}
{\cal H}_{\rm reg} & = & \case{1}{2} K \int d^2 u \sqrt{g} g^{ab}
\partial_a \tilde \theta \partial_b \tilde \theta , \nonumber \\
{\cal H}_c & = & \case{1}{2} K \int d^2 u \sqrt{g} \rho 
( - \Delta_g )^{-1} \rho ,
\end{eqnarray}
where $\rho = n - S$ is the total ``charge density" arising from
disclinations and Gaussian curvature.
Then, using $\int {\cal D} {\tilde \theta} e^{- \beta {\cal H}_{\rm reg}} = 
e^{-\beta {\cal H}_L}$ where $\beta \Ha_L = - S_L / (96 \pi)$ and $S_L$ is
the Liouville action\cite{polyakov86}
\begin{equation}
S_L = \int d^2 u \sqrt{g}R ( - \Delta_g )^{-1} R,
\end{equation} 
we can write the partition function as
\begin{eqnarray}
Z(\kappa, K , y ) & = & {\rm Tr}_{\rm v} \int {\cal D} {\bf R} 
{\cal D} {\tilde \theta} y^N e^{- \beta
{\cal H}_\kappa - \beta \Ha_\theta} \label{p-atic} \\
& = & {\rm Tr}_{\rm v} \int {\cal D} {\bf R} e^{- \beta {\cal H}_{\kappa}
- \beta {\cal H}_L - \beta \Ha_c} y^N .
\label{coulombg}
\end{eqnarray}
Here ${\cal H}_{\theta}$ and $\Ha_c$ are understood to 
depend on the total number of
vortices $N$ and on the position and charge of each vortex, ${\rm
Tr}_{\rm v}$ is the trace over vortices with approptiate factors of $N!$,
and ${\cal D}{\bf R}$ is the measure for fluid membranes, which includes the
Fadeev-Popov determinant and Liouville factor\cite{CLNP93}.  If
we include only the lowest strength vortices with $q_{\alpha} = \pm 1/p$,
then ${\rm Tr}_{\rm v}$ can be defined via
$$
{\rm Tr}_{\rm v} =  \sum_{N_+, N_-}\delta_{N_+ - N_- , p \chi }
{1 \over N_+! N_-!} \int [{\cal D} u^+]_{N_+} [{\cal D} u^-]_{N_-} ,
$$
where $N_{\pm}$ is the number of
disclinations with strength $\pm 1/p$, $\chi = 2 ( 1-h)$ 
is the Euler characteristic
of the surface, and 
$[{\cal D} u^{\pm}]_{N_{\pm}} = 
\prod_{\alpha}^{N_{\pm}} d^2 u_{\alpha}^{\pm} [g(u_{\alpha}^{\pm})]^{1/2}/a^2$
is the integration measure for the positions of the $N_{\pm}$ charges with
sign $\pm$.  
The Kroneker $\delta$ imposes the constraint that the total
disclination strength 
equal the Euler characteristic $\chi$.  On an open surface, 
the long-range Coulomb potential leads to $N_+ = N_-$, and we can use 
${\rm Tr}_{\rm v}$ with $\chi = 0$.
We can convert the Coulomb gas model [Eq. (\ref{coulombg})] to a
sine-Gordon model by representing 
the factor $e^{-\beta {\cal H}_c}$, via a
Hubbard-Stratonovich transformation, as 
$$
e^{- \beta {\cal H}_c} = e^{\beta \Ha_L}
\int {\cal D} \phi  e^{ -  \int d^2 u \sqrt{g} [{1 \over 2}(\beta K)^{-1}
\partial_a \phi \partial^a \phi - i \phi \rho] } , 
$$
where the factor $e^{\beta {\cal H}_L}$ assures that $e^{-\beta \Ha_c}$ is one 
when $\rho = 0$.
Then, setting
$\delta_{N,0} = (1/2 \pi) \int_0^{2 \pi} d\omega e^{- i
\omega  N} $, integrating over $[{\cal D} u^+]_{N_+}[{\cal D} u^-]_{N_-}$,
and shifting $\phi \to  p \phi /(2 \pi ) - \omega$, 
we obtain $Z = \int {\cal D } {\bf R} {\cal D}
\phi e^{- \cal L}$, where
\begin{eqnarray}
{\cal L} & = & \beta {\cal H}_{\kappa} + \case{1}{2} \gamma
\int d^2 u \sqrt{g} \partial_a \phi \partial^a \phi  \nonumber\\
& & -{2 y \over a^2} \int
d^2 u \sqrt{g} \cos \phi -  i \lambda  \int d^2 u \phi S ,
\label{sinegor}
\end{eqnarray}
where $\gamma = p^2/(4 \pi^2 K)$ and $\lambda = p/( 2 \pi )$.
This effective action is valid for a surface of arbitrary genus.  
The coupling between $\phi$ and geometry
is via a $\phi S$ term analagous to the dilaton
coupling of string theory\cite{dilaton}.
\par
Eqs.\ (\ref{p-atic}), (\ref{coulombg}), and (\ref{sinegor})
provide us with three equivalent versions of the partition
function for $p$-atic order on a fluctuating membrane, all of which can be
used to study the KT crinkled-to-crumpled phase transition.  We focus 
first on the Coulomb gas model on a nearly flat surface.  The Monge gauge
with ${\tilde u} = {\bf x} = ( x , y )$ and ${\bf R} ( \xv ) = (\xv , h ( \xv )) $
provides the most natural description of such a surface.  In this gauge,
the Gaussian curvature is $S = - \case{1}{2} ( \partial^2 \delta_{ij} -
\partial_i \partial_j ) \partial_i h \partial_j h$ to lowest order in $h$.
Interactions among charges in a Coulomb gas are screened by other charges.
In our Coulomb gas, there are two kinds of charges: disclination charge
and Gaussian curvature charge.  We are interested in the charge density
$\rho = n - S$.  The dielectric constant associated with this charge
density is
\begin{equation}
\epsilon^{-1} ( q ) = 
1 - (\beta K /q^2 ) C_{\rho \rho}( q ) ,
\end{equation}
where  
$$
C_{\rho\rho} ( q ) = \int d^2 x e^{-i {\bf q}\cdot ({\bf x} - \xvp)} 
\langle \sqrt{g( {\bf x} )} \rho ( \xv ) \sqrt{g(\xvp}) \rho( \xvp ) \rangle
$$
is the charge density correlation function.  This expression for
$\epsilon^{-1}$ is identical to the corresponding flat space expression
except that $\sqrt{g} \rho$ replaces $\rho$ in $C_{\rho\rho}$.  The
effective or renormalized stifness $K_R$ is related in 
the usual\cite{minnhagen} way to
the zero $q$ limit of $\epsilon^{-1}$: 
$\beta K_R = \lim_{q\rightarrow 0} \beta K \epsilon^{-1}$.
$C_{\rho\rho}(q)$ is the complete $\rho-\rho$ correlation function. It 
depends on all interactions in Eq.\ (\ref{coulombg}), 
including the Liouville term
$\beta {\cal H}_L$.  We are interested in situations in which the disclination
density, or equivalently $y$, is small, and we consider first the case
$y=0$ for which there is only Gaussian curvature charge.  Then
$C_{\rho\rho}(q) = C_{SS}^0(q)$, and $\beta {\cal H}_L + \beta \Ha_c$ is
equal to $\beta {\cal H}_c (S = 0)$ with $K$ replaced 
by $\beta K^{\prime} =  
\beta K - (12 \pi)^{-1}$ in agreement with Ref. \cite{David87-1}.  
The correlation
function $C_{SS}^0 ( q )$ can be expressed in terms of a polarization
bubble $P(q)$ as $C_{SS}^0 ( q ) = P(q) [ 1 + (\beta K^{\prime}/q^2)
P(q)]^{-1}$.  To lowest order in $(\beta \kappa)^{-1}$, $C_{SS}^0(q) = P(q)$
with 
\begin{equation}
P( {\bf q} ) = {1 \over 2} {1 \over \beta^2 \kappa^2} \int {d^2 k \over (2 \pi
)^2} {|{\bf q} \times {\bf k}|^4 \over k^4 |\kv + \qv |^4} = {3 \over 32 \pi} {q^2
\over \beta^2 \kappa^2} .
\end{equation}
Thus, in the absence of
disclinations, the renormalized rigidity is 
$\beta{\overline K} = \beta K -(3/32
\pi) (K/\kappa)^2$.  It is straightforward to verify that the correlation
$g( {\bf x} ) = \langle \cos ( \theta( \xv ) - \theta ( 0 )  - \int_0^{\xv}
ds_a A^a) \rangle$ between a spin at the origin and a spin at ${\bf x}$
parallel transported to the origin decays with $|{\bf x}|$ with an exponent
determined by $K_R$ rather than $K$: $g( {\bf x} ) \sim
|{\bf x}|^{- 2 \pi \beta K_R}$, which reduces when $y=0$ to $|\xv|^{- 2 \pi
\beta {\overline K}}$.
\par
Next we consider the order $y^2$ terms arising from disclinations.  To
lowest order in $\beta \kappa$, the disclination density correlation
function $C_{nn} (q)$ has the same form as it has in flat space:
$C^0_{nn} ( q ) =  q^2 4 \pi^3 y^2 \int_a^{\infty} 
dr r^{3 - 2 \pi \beta K}$, where $a$ is the short distance cutoff.  The cross
correlation function $C_{nS}(q)$ is also of order $y^2$ and equal to
$\beta K C_{SS}^0 (q ) C_{nn}^0 ( q )/q^2$.  An expansion in powers of
$\beta K/\beta \kappa$ then yields
\begin{equation}
(\beta K_R)^{-1} = (\beta {\overline K})^{-1} + 
4 \pi^3 y^2 \int_a^{\infty} {dr
\over a} \left({r \over a}\right)^{3 - 2 \pi \beta K_R}.
\end{equation}
This is identical to the equation
for $K_R$ on a rigid flat surface\cite{KT} 
but with $(\beta {\overline K})^{-1} = (\beta K)^{-1} +
(3/32 \pi)(\beta\kappa)^{-2}$ replacing $(\beta K)^{-1}$.
It yields the KT recursion relations for 
$\beta {\overline K}$ and $y$. To these we
must add the equation governing the renormalization of $\kappa$. 
To lowest order in the temperture $\beta^{-1}$ and $y^2$, the equation
governing $\kappa$ is identical to that derived in Ref. \cite{David87-1}.
To leading order in $\beta^{-1}$ and $y$, we can replace $K$ 
by ${\overline K}(l)$ in the equation for $\kappa$.
Our recursion relations to lowest order in $\beta^{-1}$ and $y$  are
\begin{eqnarray}
{d (\beta {\overline K})^{-1} \over dl} 
&= &{4 \pi ^2 \over p^2} y^2 
\label{recur1}\\
{dy \over dl}& =& \left( 2 - 
{\pi \beta {\overline K} \over p^2} \right) y
\label{recur2} \\
{d \beta \kappa \over dl} &= & - {3 \over 4 \pi } \left ( 1 - {\beta
{\overline K} \over 4 \beta \kappa} \right ) .
\label{recur3}
\end{eqnarray}
We have also derived these equations directly from the sine-Gordon model
using the procedure or Ref.\ \onlinecite{AmitGoldGrin80}.
They are what one would have expected.  The first two equations 
are exactly the KT equations, but with
$\beta K$ replaced by $\beta {\overline K}$.  Similarly, the equation
controling $\beta \kappa$ is identical to that derived using a momentum
shell renormalization procedure in Refs.\ \onlinecite{PelitiLeibler85-1} 
but again with
$\beta {\overline K}$ replacing $\beta K$.  
Eqs.\ (\ref{recur1}) to (\ref{recur3}) 
have a fixed line corresponding to the crinkled phase at
$y=0$, $\beta \kappa = \beta {\overline K} /4$ independent of $p$.
At $\beta {\overline K} = \beta
{\overline K}^*= 2/\pi$, the system becomes unstable with respect to the
formation of disclinations, and there is a KT transition to a fluid phase
with unbound disclinations.  For $(\beta {\overline K})^{-1}> (\beta
{\overline K}^*)^{-1}$, $\beta {\overline K}$ decreases with $l$, and
the ratio $\beta {\overline K} / \beta \kappa$ tends to zero so that
Eq.\ (\ref{recur3}) for $\beta \kappa$ tends to that for a free fluid
membrane.  An altenative route to the derivation of Eqs. (\ref{recur1}) and
(\ref{recur2}) is to integrate out height fluctuations to produce an
effective theory for $\theta$ (or $\phi$ in the sine-Gordon model) on a
coarse-grained flat surface.  This effective theory is identical to the
usual flat space theory with ${\overline K}$ replacing $K$.    
\par
\begin{figure}[h]
\centerline{\psfig{figure=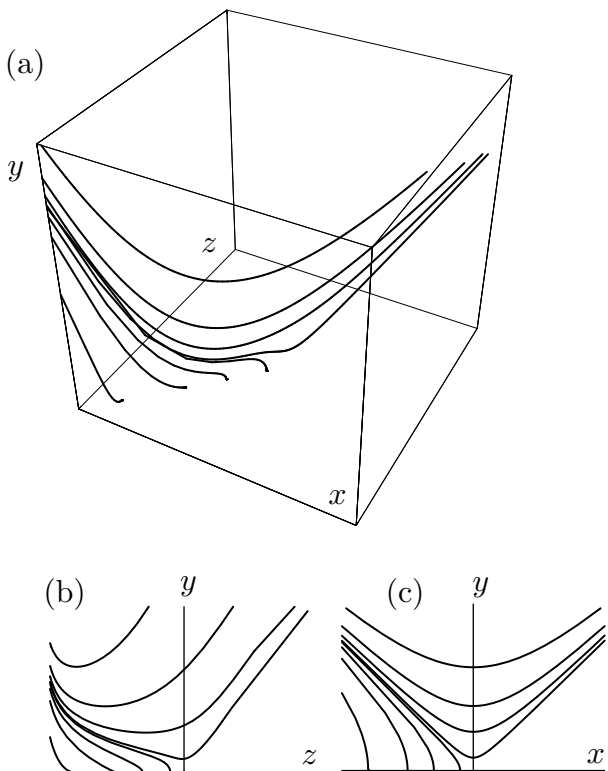}}
\caption{Renormalization flows Eqs.\ (\protect\ref{nrecur1}) and
(\protect\ref{nrecur3}).  (a) shows three-dimensional flows in $x$, $y$, and $z$.
(b) and (c) show, respectively, flows projected onto the $(y,z)$ and $(y,x)$
planes, which are similar to those in the $(y,x)$-plane of a rigid flat
surface.}
\label{fig2}
\end{figure}

In the vicinity of the KT critical point, $y= 0$, 
$\beta {\overline K}=
\beta {\overline K}^*$, $\beta \kappa = \beta \kappa^* \equiv \beta
{\overline K}^*/4$, we can set $\beta {\overline K} = \beta {\overline
K}^* ( 1 - x )$, and $\beta \kappa = \beta \kappa^* ( 1 - z )$.  To
lowest nontrivial order in $x,y,z$, Eqs.\ (\ref{recur1}) to (\ref{recur3})
become
\begin{equation}
{d x \over dl } =  8 \pi^2 y ^2 , \qquad
{d y \over dl} = 2 xy 
\label{nrecur1}
\end{equation}
\begin{equation}
{d z \over dl }=  {3 \over 2} [ x - ( 1-x ) z ]
\label{nrecur3}
\end{equation}
The flow lines for these equations are shown in Figs.\ \ref{fig1} and
\ref{fig2}.  In the $xy$ plane (Fig.\ \ref{fig2}c), 
flows are identical to the KT flows.  In the $xz$ 
plane (Fig.\ \ref{fig2}a), there is
some curvature in the flows toward the fixed line resulting from couplings
to $y$.  In addition, flows are away from the critical line $x=0$.  Neither
of these effects was observed in previous treatments ignoring
$y$\cite{GuitKardar90-1}.  Fig.\ \ref{fig1}b shows flows in the 
$(\beta K)^{-1} - (\beta \kappa )^{-1}$ rather than the $(\beta
{\overline K})^{-1} - (\beta \kappa )^{-1}$ plane.  This shows that the
crinkled phase exists only in a region near the origin with both $(\beta
K)^{-1}$ and $(\beta \kappa )^{-1}$ small.  Thus, decreasing the bending
rigidity $\kappa$ will lead to disordering to the fluid phase.  This effect
is not apparent in the treatment of Ref.\ \onlinecite{GuitKardar90-1}. 
\par
The persistance length in the fluid phase can be obtained with the aid of
Eqs.\ (\ref{nrecur1}) to (\ref{nrecur3}).  
$\beta {\overline K}$ reaches $0$ when $x=1$, and
does not become negative.  We may, therefore, use Eq.\ (\ref{nrecur1})
to determine $x(l)$ for $l < l_0$ where $x(l_0) = 1$.
The KT correlation length is $\xi_{KT} = a e^{l_0} = a \exp [\pi/b
\sqrt{T-T_c}]$.  For $l> l_0$, we set $x(l) =1$.  Thus $z(l)$ satisfies
Eq.\ (\ref{nrecur3}) for $l< l_0$ and $dz/dl = 3/2$ for $l> l_0$, 
and
\begin{equation}
z( l ) = \cases{z_0 + F(l), & if $l\leq l_0$\cr
                z_0 + {3 \over 2} ( l - l_0 ) + F( l_0 ) & if $l > l_0$ ,\cr}
\end{equation}
where $z_0 = 1 -(\beta \kappa/\beta \kappa^*)$ and
$F(l) = \int_0^l d l e^{- \psi ( l ) } x(l)$ with
$\psi ( l ) = (3/2) \int _0^l dl ( 1 - x(l) )$.  The persistence length is
then $\xi_p = a e^{l^*}$ where $z( l^* ) = 1 $.  For large $\kappa$, $l^* >
l_0$, and $\xi_p = a e^{F(l_0)} \xi_{KT} e^{4 \pi \kappa / 3 T}$.  For
smaller $\kappa$ and for $\beta {\overline K}$ near the critical point, 
$l^*$ may be less than $l_0$ and $\xi_p$ can be less than $\xi_{KT}$.    
\par
The sine-Gordon model of Eq.\ (\ref{sinegor}) applies to surfaces of arbitrary
genus.  It can be used to investigate the effect of nonzero Gaussian
curvature on the KT transition.  
On a nearly spherical surface of genus zero, we can parametrize
the surface by its radius vector as a function of solid angle $\Omega$:
${\bf R} ( \Omega ) = R_0 ( 1 + \eta ( \Omega ) ) {\bf e}_r$,
where ${\bf e}_r$ is the unit radius vector.  To lowest order
in $\eta$,
\begin{eqnarray}
{\cal L} & = & - \case{1}{2} \beta \kappa \int d \Omega ( (\nabla^2 + 2) \eta
)^2 + \case{1}{2} \gamma \int d\Omega 
( {\mbox{\boldmath{$\nabla$}}} \phi )^2 \nonumber \\
& & - i \lambda \int d\Omega \phi ( \nabla^2 + 2 ) \eta + {2 y \over (a/R_0)^2}
\int d\Omega \cos \phi ,
\end{eqnarray}
where ${\mbox{\boldmath{$\nabla$}}}$ is the gradient 
on a spherical surface of unit radius.
Note that there is a linear coupling between $\phi$ and $\eta$.  In the
case of a nearly flat membrane, this coupling was 
proportional to $\phi ( \nabla^2 h)^2$ and thus of higher order in normal
displacments of the surface.  
An effective theory on the reference
spherical surface can be obtained by integrating out the height
fluctuations.  The resulting action is
$$
{\cal L} = \case{1}{2} \gamma \int d\Omega \phi [ - \nabla^2  + (K/\kappa) ]
\phi +{2 y \over (a/R_0)^2} \int d\Omega \cos \phi .
$$
The important feature of this model is the ``mass term", $(K/\kappa)
\phi^2$ that supresses fluctuations in $\phi$ and keeps the system in the
ordered phase of the sine-Gordon model or, alternatively, in the
disordered phase of the Coulomb gas model.  If $K/\kappa > 1$, the
transition is supressed altogether, much as the KT transition is supressed
in an infinite 2D superconductor.  If on the other hand $K/\kappa \ll 1$,
there can be an effective transition.  In particular, when $\kappa
\rightarrow \infty$, the KT transition on a rigid 
sphere\cite{ovrutthomas91} is regained.
\par
This work was supported in part by the Penn Laboratory for Research in the
Structure of Matter under NSF grant No.  DMR 91-20668.  TCL is also grateful for
support under NSF grant No. PHY 89-04035 
from the Institute for Theoretical Physics, where a portion of this 
was carried out.  We are grateful to Leo Radzihovsky and Mark Bowick for helpful
conversations.

\bibliographystyle{prsty}

\begin{thebibliography}{10}

\bibitem{Helfrich73-1}
W. Helfrich, Z. Naturforsch. {\bf 28C},  693  (1973);
P. Canham, J. Theor. Biol. {\bf 26},  61  (1970).

\bibitem{Jerusalem89}
{\em Statistical Mechanics of Membranes and Surfaces}, edited by D. Nelson, T.
Piran, and S. Weinberg (World Scientific, Singapore, 1989).

\bibitem{Helfrich85-1}
W. Helfrich, J. Phys. (Paris) {\bf 46},  1263  (1985);
J. Phys. (Paris) {\bf 47},  322  (1986).

\bibitem{PelitiLeibler85-1}
L. Peliti and S. Leibler, Phys. Rev. Lett. {\bf 54},  1690  (1985).

\bibitem{Nelson87-1}
D. R. Nelson and L. Peliti, J. Phys. (Paris) {\bf 48},  1085  (1987).

\bibitem{Nelson79-1}
D.~R. Nelson and B. Halperin, Phys. Rev.B {\bf 19},  2475  (1979).

\bibitem{David87-1}
F. David, E. Guitter, and L. Peliti, J. Phys. {\bf 49},  2059  (1987).

\bibitem{GuitKardar90-1}
E. Guitter and M. Kardar, Europhys. Lett. {\bf 13},  441  (1990).

\bibitem{dilaton}
See for example, M.B. Green, J.H. Schwarz, and E. Witten, {\it Superstring
Theories} (Cambridge University Press, Cambridge, 1987).

\bibitem{AmitGoldGrin80}
D. Amit, Y. Goldschmidt, and G. Grinstein, J. Phys. A {\bf 13},  585  (1980).

\bibitem{polyakov86}
A. Polykov, Nucl. Phys. {\bf B268},  406  (1986).

\bibitem{CLNP93}
W. Cai, T. Lubensky, T. Powers, and P. Nelson, J. Phys. II (France) {\bf 4},  931
   (1994).

\bibitem{minnhagen}
Peter Minnhagen, Revl Mod. Phys. {\bf 59}, 1001 (1987).

\bibitem{KT}
J.M. Kosterlitz and D.J. Thouless, J. Phys. C{\bf 5}, L124 (1972);
J. Jos\'{e}, L.P. Kadanoff, S. Kirkpatrick, and D.R. Nelson, Phys. Rev. 
B{\bf16}, 1217 (1977);E Phys. Rev. B{\bf 17}, 1477 (1978).

\bibitem{ovrutthomas91}
B.~A. Ovrut and S. Thomas, Phys. Rev. D {\bf 43},  1314  (1991).

\end{thebibliography}

\end{document}